\documentstyle[preprint,prb,aps]{revtex}

\begin{document}
\title{Interference Effects between Three Coupled Bose-Einstein Condensates}
\author{Sun Zhang and Fan Wang}
\address{Department of Physics, Nanjing University, Nanjing, 210093, China}
\maketitle

\begin{abstract}
We study the interference effects between three weakly linked trapped
Bose-Einstein condensates (BEC) as a generalization of the two-component
condensates. Three coupled Gross-Pitaevskii equations (GPE) are used to
describe the dynamics of the system. The nonsinusoidal oscillation is found
as a generalization of the Josephson effect in superconductivity. The
self-trapped effects are also predicted in three coupled BEC. Moreover, in
general case, the phase diagrams of the system are closed only for some
special parameters, which can be used to determine the interaction
parameters between atoms in BEC.
\end{abstract}


Recently, Bose-Einstein Condensate (BEC) has been realized in a dilute and
ultracold gas of trapped alkali-metal atoms \cite{1,2,3}. It opens a new
field for studying the physics of BEC, which is hoped to improve our
understanding on the fundamental concept of quantum mechanics and quantum
many body problem \cite{4,5}. From the ensuing theoretical and experimental
works, it is clear that the condensed Bose gas displays the crucial
properties of a confined coherent matter \cite{5}.

Since the current understanding of BEC is mainly based on the mean-field
approximation, where a macroscopic wave function is introduced as the order
parameter, the study of this feature by investigating the interference
phenomena should be of great importance \cite
{5,6,7,8,9,10,11,12,13,14,15,16,17,18,19}. Such a description using a
macroscopic wave function with a definite phase implies U(1) gauge symmetry
broken.

There are many methods suggested theoretically for detecting the
interference effects between two components of condensates. The one is the
investigation using the continuous measurement theory \cite{6}. It shows
that an interference pattern between two condensates can be built up
dynamically in a single run of an experiment, even though no phases have
even been assumed \cite{12,13,14}. Another approach is that, one can set up
a potential barrier between two trapped condensates by using a far
off-resonant laser. If one lowers the potential barrier produced by the
laser, the atoms would tunnel through the barrier due to weak coupling
between these two condensates. The interference is developed, somewhat like
Josephson effect in superconductivity \cite{15,16}. Similarly, if one
switches off the laser potential traps, the atoms would expanded
ballistically due to the sudden disappearance of the barrier. The
interference fringes can be observed \cite{17,18}. It is also interesting
experimentally that the interference effects can also be tested by means of
different hyperfine atomic states in a single trap \cite{19}.

Among these methods mentioned above, some have been realized in experiments 
\cite{20,21,22,23}. In Ref. [20], the authors reported that fringes were
observed in the density of two overlapping condensates, released after
switching off the traps. In Ref. [21], the authors measured the relative
phase of two condensates in different hyperfine atomic states occupying a
single trap, using Ramsey's method of separated oscillating fields \cite{24}%
. From experimental data \cite{18,20}, the phase locking indeed occurs for
small separation between condensates, implying the broken gauge symmetry.

The interference effects of two component condensates have been studied in a
variety of cases, but there are very few on three \cite{5}. In this paper,
we would discuss the interference effects between three coupled BEC.

Any generalization on this topic would increases the difficulty
significantly. The reward is that the more complex nonlinearity and the more
parameters in the basic equations would bring richer dynamics and more
compound structure into the model. Experimentally, a far off-resonant laser
barrier can be used to split a trapped condensate into two, three or even
more well-trapped condensates. Lowering the laser intensity allows atoms
tunneling through the barrier. Then interference will be built up between
these condensates. The optical confinement method also has the possibility
to create a three component BEC in different hyperfine atomic states \cite
{25}.

Due to the U(1) gauge symmetry broken, it is crucial to introduce a global
phase to the order parameter, the macroscopic wave function $\Psi ({\bf r},t)
$ to describe BEC \cite{15,16,17,18,26}. Then, the wave function $\Psi ({\bf %
r},t)$ at $T=0$ can be factorized as \cite{15,16,26} 
\begin{equation}
\Psi ({\bf r},t)=\psi (t)\Phi ({\bf r}),  \label{1}
\end{equation}
which satisfies a nonlinear Shr\"{o}dinger equation, the Gross-Pitaevskii
equation (GPE) \cite{27}, 
\begin{equation}
i\hbar \frac{\partial \Psi ({\bf r},t)}{\partial t}=-\frac{\hbar ^2}{2m}%
\nabla ^2\Psi ({\bf r},t)+[V_{trap}({\bf r})+g_0\left| \Psi ({\bf r}%
,t)\right| ^2]\Psi ({\bf r},t),  \label{2}
\end{equation}
where $V_{trap}({\bf r})$ is the external trap potential and $g_0=4\pi \hbar
^2a/m$, $m$ the atomic mass and $a$ the scattering length of the atom-atom
interaction. When we study the interference of three weakly coupled BEC in
traps 1,2 and 3, the dynamics of the system is described by three GPE,
coupled with tunneling matrix elements. It can be written as a three-state
model \cite{28} 
\begin{eqnarray}
i\hbar \frac{\partial \psi _1}{\partial t} &=&(E_1^0+U_1N_1)\psi _1-K\psi _2,
\label{3aa} \\
i\hbar \frac{\partial \psi _2}{\partial t} &=&(E_2^0+U_2N_2)\psi _2-K\psi
_1-K\psi _3,  \label{3bb} \\
i\hbar \frac{\partial \psi _3}{\partial t} &=&(E_3^0+U_3N_3)\psi _3-K\psi _2,
\label{3cc}
\end{eqnarray}
with 
\begin{equation}
\psi _{1,2,3}=\sqrt{N_{1,2,3}}\exp (i\theta _{1,2,3}(t)),  \label{4a}
\end{equation}
where $N_{1,2,3}$ and $\theta _{1,2,3}$ are the atom numbers and
(time-dependent) phases for BEC in trap 1, 2 and 3, respectively. $K$ is the
coupling matrix element between trap 1 and 2 or trap 2 and 3 (for
simplicity, we only consider the case, that these two coupling matrix
elements are the same and there is no coupling between 1 and 3). $E_{1,2,3}^0
$ are the zero-point energies in each well, $U_{1,2,3}N_{1,2,3}$ are
proportional to the atomic self-interaction energies, induced by the overlap
of the spatial part of the macroscopic wave function, and the total number
of atoms $N_T=N_1+N_2+N_3$ is a constant.

From Eqs. (\ref{3aa}), (\ref{3bb}), (\ref{3cc}) and (\ref{4a}), we get four
independent equations in terms of the phase difference and the atom
population imbalance 
\begin{eqnarray}
\hbar \frac{\partial \phi _1}{\partial t} &=&(E_1^0-E_2^0)+(U_1N_1-U_2N_2)+K%
\frac{N_1-N_2}{\sqrt{N_1N_2}}\cos \phi _1+K\sqrt{\frac{N_3}{N_2}}\cos (\phi
_1-\phi _2),  \label{5a} \\
\hbar \frac{\partial \phi _2}{\partial t} &=&(E_1^0-E_3^0)+(U_1N_1-U_3N_3)-K%
\sqrt{\frac{N_2}{N_1}}\cos \phi _1+K\sqrt{\frac{N_2}{N_3}}\cos (\phi _1-\phi
_2),  \label{5b}
\end{eqnarray}
\begin{equation}
\frac \hbar 2\frac{\partial (N_1-N_2)}{\partial t}=-2K\sqrt{N_1N_2}\sin \phi
_1-K\sqrt{N_2N_3}\sin (\phi _1-\phi _2),  \label{5c}
\end{equation}
\begin{equation}
\frac \hbar 2\frac{\partial (N_1-N_3)}{\partial t}=-K\sqrt{N_1N_2}\sin \phi
_1+K\sqrt{N_2N_3}\sin (\phi _1-\phi _2),  \label{5d}
\end{equation}
where $\phi _1=\theta _2-\theta _1$ and $\phi _2=\theta _3-\theta _1$.

In order to get an understanding on the novel phenomena of interference
effects between three trapped BEC, let us consider an ideal case: the
totally symmetric case, in which, $U_1=U_2=U_3\equiv U$, $E_1^0=E_2^0=E_3^0$%
, and the atom population imbalances of traps 1, 2 and of traps 2, 3 are the
same, that is $N_1-N_2=N_3-N_2$, or $N_1=N_3$. Then, we can simplify Eqs. (%
\ref{5a}), (\ref{5b}), (\ref{5c}) and (\ref{5d}) as 
\begin{equation}
\hbar \frac{\partial \phi _1}{\partial t}=U(N_1-N_2)+K\frac{N_1-N_2}{\sqrt{%
N_1N_2}}\cos \phi _1+K\sqrt{\frac{N_1}{N_2}}\cos \phi _1,  \label{6a}
\end{equation}
\begin{equation}
\frac \hbar 2\frac{\partial (N_1-N_2)}{\partial t}=-3K\sqrt{N_1N_2}\sin \phi
_1.  \label{6b}
\end{equation}
After rescaling the time variable and the population imbalance, $2Kt/\hbar
\rightarrow t$, $z\equiv \frac{N_1-N_2}{N_T}$, Eqs. (\ref{6a}) and (\ref{6b}%
) can be written as 
\begin{eqnarray}
\frac{\partial z}{\partial t} &=&-\sqrt{1-z-2z^2}\sin \phi ,  \label{7a} \\
\frac{\partial \phi }{\partial t} &=&\Lambda z+\frac 12\frac{4z+1}{\sqrt{%
1-z-2z^2}}\cos \phi ,  \label{7b}
\end{eqnarray}
where $\Lambda \equiv \frac{UN_T}{2K}$, $\phi \equiv \phi _1$.

If we treat $z(t)$ and $\phi (t)$ as a pair of canonically conjugate
variables, it is not difficult to construct a Hamiltonian 
\begin{equation}
H=\frac \Lambda 2z^2-\sqrt{1-z-2z^2}\cos \phi ,  \label{8}
\end{equation}
satisfying the canonical equations, 
\begin{eqnarray}
\stackrel{.}{z} &=&-\frac{\partial H}{\partial \phi },  \label{9a} \\
\stackrel{.}{\phi } &=&\frac{\partial H}{\partial z}.  \label{9b}
\end{eqnarray}
In Eq. (\ref{8}), $z<0.5$, consistent with the totally symmetric conditions.
The fluctuations of the phases and the atom numbers must be small and has
not been included here \cite{15}.

Now, let us discuss Eqs. (\ref{7a}) and (\ref{7b}) in more details.

(1) {\it zero-phase mode}. With initial conditions $z(0)=0.3$ and $\phi
(0)=0 $, we can solve Eqs. (\ref{7a}) and (\ref{7b}) numerically in
zero-phase mode. Some results are given in Fig. \ref{1}. $\Lambda $ takes on
the values 10, 20, 38, 38.25 and 39 for Fig. 1(a), 1(b) ,1(c), 1(d) and
1(e), respectively. The anharmonic nonsinusoidal oscillations are the
generalized sinusoidal Josephson oscillations \cite{29}. $z$ is asymmetric
about its zero point because of the existence of the third trap, $%
\left\langle z(t)\right\rangle \neq 0$. As $\Lambda $ increases, the
oscillations around $z=0$ become anharmonic [Figs. 1(a), 1(b) and 1(c)].
When $\Lambda $ exceeds a critical value $\Lambda _c=38.25$, the populations
oscillates around some nonzero values, and $z(t)>0$ [Fig. 1(e)]. This
population imbalance self-trapped effect is somewhat like the pendulum bob
swing over the $\phi =\pi $ vertical orientation for sufficiently large
initial values. This state can be achieved from different approaches. In
zero-phase mode, there is a pair of eigenfunctions of GPE, $z=0$ and $\phi
=\pi $, and the ground state energy is $E=1$. If the self-trapped effect
takes place, $z\neq 0$ when $\phi =\pi $, then 
\begin{equation}
H(z(0),\phi (0))=\frac \Lambda 2z(0)^2-\sqrt{1-z(0)-2z(0)^2}\cos \phi (0)>1,
\label{10}
\end{equation}
so the critical values depend on the following condition 
\begin{equation}
\Lambda _c=2(\frac{1+\sqrt{1-z(0)-2z(0)^2}\cos \phi (0)}{z(0)^2}).
\label{11}
\end{equation}
Here, $z(0)=0.3$ and $\phi (0)=0$, so $\Lambda _c=38.25$.

(2) $\pi ${\it -phase mode}. If we select the initial conditions as $\phi
(0)=\pi $ and $z(0)=0.3$, we can discuss the interference dynamics of $\pi $%
-phase oscillations from Eqs. (\ref{7a}) and (\ref{7b}). The numerical
results are shown in Fig. \ref{2}. It is clear that the nonsinusoidal
oscillations in $\pi $-phase mode are similar to that in zero-phase mode
[Fig. 2(a)], but the self-trapped effects are not. There are two kinds of
self-trapped effects in $\pi $-phase mode $\left\langle z(t)\right\rangle
<z(0)$ and $\left\langle z(t)\right\rangle >z(0)$ as given in Figs. 2(d) and
2(f). According to the degenerate GPE eigenstates that break the $z$
symmetry, we can get the critical value of $\Lambda _c$ for these two
different self-trapped effects 
\begin{equation}
\Lambda _c=\frac{4z(0)+1}{2z(0)\sqrt{1-z(0)-2z(0)^2}},  \label{12}
\end{equation}
here, $z(0)=0.3$, so $\Lambda _c=5.08$. If $\Lambda $ exceeds the value of $%
\Lambda _c$, the system goes from the first kind of self-trapped state [Fig.
2(d)] into the second kind of self-trapped state [Fig. 2(f)].

(3) $z$-$\phi $ {\it phase diagrams}. $z$ and $\phi $ are two canonically
conjugate variables in Eqs. (\ref{7a}) and (\ref{7b}), their dynamical
behavior in $z$-$\phi $ phase diagrams are shown in Fig. \ref{3}. In Figure
3(a), $\Lambda =10$ and $\phi (0)=0$, it is the zero-phase mode. For $%
z(0)=0.20$, and $0.40$, the trajectories are closed, asymmetric curves,
corresponding to the generalized nonsinusoidal oscillations. For $z(0)=0.487$%
, it is the critical case (here, for the zero-phase mode, we should solve
the critical value of $z$ from Eq. (\ref{11}) with $\Lambda =10$). And for $%
z(0)=0.498$, the running self-trapped case, $z(t)$ is locked but $\phi (t)$
is unlocked, $-\infty <\phi (t)<\infty $. In Fig. 3(b), $\Lambda =7$ and $%
\phi (0)=\pi $, the $\pi $-phase mode, the system is self-trapped for all
values of $z(0)$. For $z(0)=0.08$, it is the running-mode self-trapped case, 
$\phi $ is unbounded, $-\infty <\phi (t)<\infty $, while for larger $z(0)$,
above a critical value of $z_c=0.111$ according to Eq. (\ref{12}) (here, for
the $\pi $-phase mode, we should solve the critical value of $z$ from Eq. (%
\ref{12}) with $\Lambda =7$), $\phi (t)$ is also localized around $\pi $. So
in Fig. 3(b), there are two kinds of self-trapped effects for $z(0)=0.08$
and for $z(0)=0.111$, $0.30$ and $0.37$ in $\pi $-phase mode, respectively.

The self-trapped effects in coupled BEC are the results of the nonlinearity
of interatomic interaction in GPE and the long-range quantum coherence of a
macroscopic number of atoms.

(4) $z$-$\phi $ {\it phase diagrams, the general case. }From Eqs. (\ref{5a}%
), (\ref{5b}), (\ref{5c}) and (\ref{5d}) generally, after rescaling the atom
population imbalance variables $z_1\equiv \frac{N_1-N_2}{N_T}$ and $%
z_2\equiv \frac{N_1-N_3}{N_T}$, and time $2Kt/\hbar \rightarrow t$, we can
get four coupled dynamical equations 
\begin{eqnarray}
\frac{\partial \phi _1}{\partial t} &=&\frac 12\frac{3z_1}{\sqrt{%
(z_1+z_2+1)(z_2+1-2z_1)}}\cos \phi _1+\frac 12\sqrt{\frac{z_1+1-2z_2}{%
z_2+1-2z_1}}\cos (\phi _1-\phi _2)  \nonumber \\
&&+\frac 13\Lambda (z_1+z_2+1)-\frac 13\Lambda (z_2+1-2z_1),  \label{20a} \\
\frac{\partial \phi _2}{\partial t} &=&-\frac 12\sqrt{\frac{z_2+1-2z_1}{%
z_1+z_2+1}}\cos \phi _1+\frac 12\sqrt{\frac{z_2+1-2z_1}{z1+1-2z_2}}\cos
(\phi _1-\phi _2)  \nonumber \\
&&+\frac 13\Lambda (z_1+z_2+1)-\frac 13\Lambda (z_1+1-2z_2),  \label{20b} \\
\frac{\partial z_1}{\partial t} &=&-\frac 23\sqrt{(z_1+z_2+1)(z_2+1-2z_1)}%
\sin \phi _1  \nonumber \\
&&-\frac 13\sqrt{(z_2+1-2z_1)(z_1+1-2z_2)}\sin (\phi _1-\phi _2),
\label{20c} \\
\frac{\partial z_2}{\partial t} &=&-\frac 13\sqrt{(z_1+z_2+1)(z_2+1-2z_1)}%
\sin \phi _1  \nonumber \\
&&+\frac 13\sqrt{(z_2+1-2z_1)(z_1+1-2z_2)}\sin (\phi _1-\phi _2).
\label{20d}
\end{eqnarray}
Here, for simplicity, symmetric parameters are also considered, $%
U_1=U_2=U_3\equiv U$, $E_1^0=E_2^0=E_3^0$, and $\Lambda \equiv \frac{UN_T}{2K%
}$. Besides the nonsinusoidal oscillations and self-trapped effects, we can
find another novel phenomenon between three trapped BEC. With some initial
conditions ($\phi _1(0)$, $\phi _2(0)$, $z_1(0)$, $z_2(0)$ and $\Lambda $),
the trajectory of $z$-$\phi $ phase-diagrams ($z_1$-$\phi _1$, $z_2$-$\phi
_2 $) are not closed. They are unclosed for general initial conditions, but
closed for some special initial values. As shown in Fig. \ref{4}, we fix the
values of $\phi _1(0)=0$, $\phi _2(0)=0,$ $z_1(0)=0.3$ and $z_2(0)=0.1$, and
let $\Lambda $ varied. When $\Lambda =2$ [Fig. 4(a) and 4(b)] and $\Lambda
=9 $ [Fig. 4(e) and 4(f)], the $z$-$\phi $ phase-diagrams are unclosed. But
for $\Lambda =6.8$ [Fig. 4(c) and 4(d)], they are convergent and closed.
These phase-diagrams exhibit a transition from unclosure to closure, then to
unclosure again during changing the values of $\Lambda $. Only a special
value of $\Lambda $ can make phase-diagrams close under some values of $\phi
_1(0)$, $\phi _3(0),$ $z_1(0)$ and $z_2(0)$. This property can be used to
determine the value of $\Lambda $ in experiments.

This feature of phase-diagrams originates from two aspects. One is the
nonlinearity of the interatomic interaction. It is known that the
phase-diagram of a nonlinearity system is always not closed \cite{30}. Eqs. (%
\ref{20a}), (\ref{20b}), (\ref{20c}) and (\ref{20d}) are four coupled
nonlinear equations, so the phase-diagrams are generally unclosed. The other
is the interaction between three trapped BEC. The interference of BEC in a
macroscopic number of atoms determine the properties of phase-diagrams. The
detail study on this complicated problem would be a special topic elsewhere 
\cite{31}.

In summary, the interference of three trapped BEC induces population
nonsinusoidal oscillations which is a generalization of sinusoidal Josephson
effect in superconductivity. The quantum self-trapped effects occur in three
BEC for both zero-phase mode (one kind of self-trapped state) and $\pi $%
-phase mode (two kinds of self-trapped states) when parameters exceed some
critical values. Any observation of the predicted interference effects would
unambiguously prove the existence of gauge symmetry broken and the relative
phase between condensates. This consideration, concerning the interference
between three Bose condensates, allows the numerical simulation of some
recent experiments \cite{25}. From calculations, we have inferred that the
convergence of the phase diagrams is sensitive to the adjustment of the
system parameters. The phase-diagrams are not closed in general cases, but
for some special conditions, the phase-diagrams are convergent and closed,
which can be used to determine the interaction parameter between atoms of
BEC in experiments.

\begin{figure}[tbp]
\caption{Rescaled population imbalance $z(t)$ versus dimensionless time
variable $t$, with initial conditions $z(0)=0.3$ and $\phi (0)=0$ in
zero-phase mode. The interaction parameter $\Lambda $ takes the values (a)
10, (b) 20, (c) 38, (d) 38.25 and (e) 39.}
\label{1}
\end{figure}

\begin{figure}[tbp]
\caption{Rescaled population imbalance $z(t)$ versus dimensionless time
variable $t$, with initial conditions $z(0)=0.3$ and $\phi (0)=\pi $ in $\pi 
$-phase mode. The interaction parameter $\Lambda $ takes the values (a) 2,
(b) 4, (c) 5, (d) 5.05 (e) 5.08 and (f) 5.1.}
\label{2}
\end{figure}

\begin{figure}[tbp]
\caption{Phase diagrams of population imbalance $z$ versus phase difference $%
\phi $. (a) and (b) are corresponding to zero-phase mode $\phi (0)=0$ with $%
\Lambda =10$ and $\pi $-phase mode $\phi (0)=\pi $ with $\Lambda =7$,
respectively. The values of $z(0)$ are as given.}
\label{3}
\end{figure}

\begin{figure}[tbp]
\caption{Phase diagrams of population imbalance versus phase difference in
general case, with $\phi _1(0)=0$, $\phi _2(0)=0,$ $z_1(0)=0.3$ and $%
z_2(0)=0.1$. (a), (c) and (e) for $z_1-\phi _1$ phase diagrams, while (b),
(d) and (f) for $z_2-\phi _2$ phase diagrams. The dimensionless time $%
-65<t<65$. $\Lambda $ takes the values 2 for (a) and (b), 6.8 for (c) and
(d), 9 for (e) and (f).}
\label{4}
\end{figure}

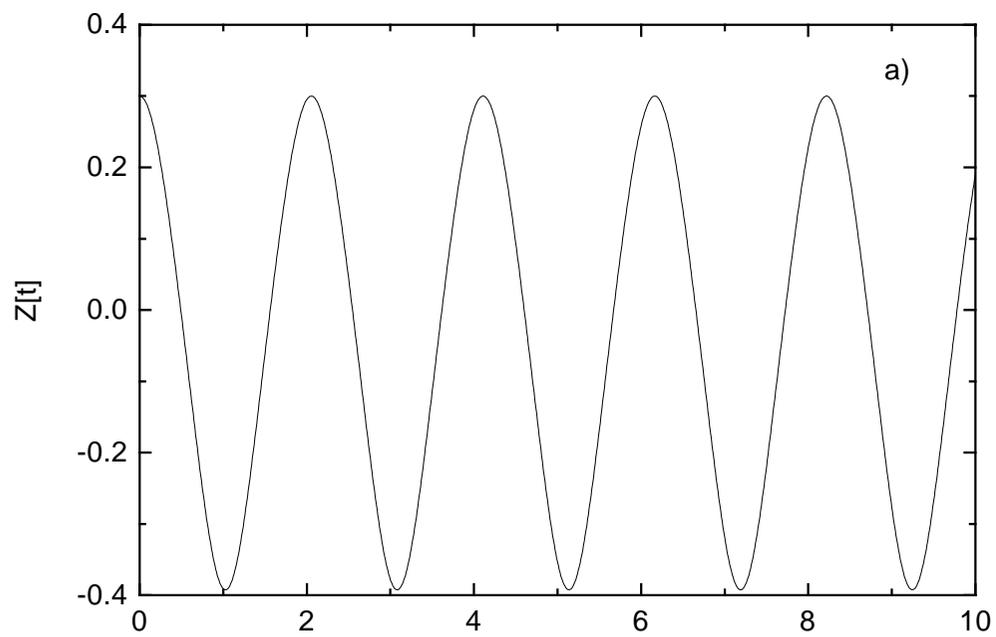
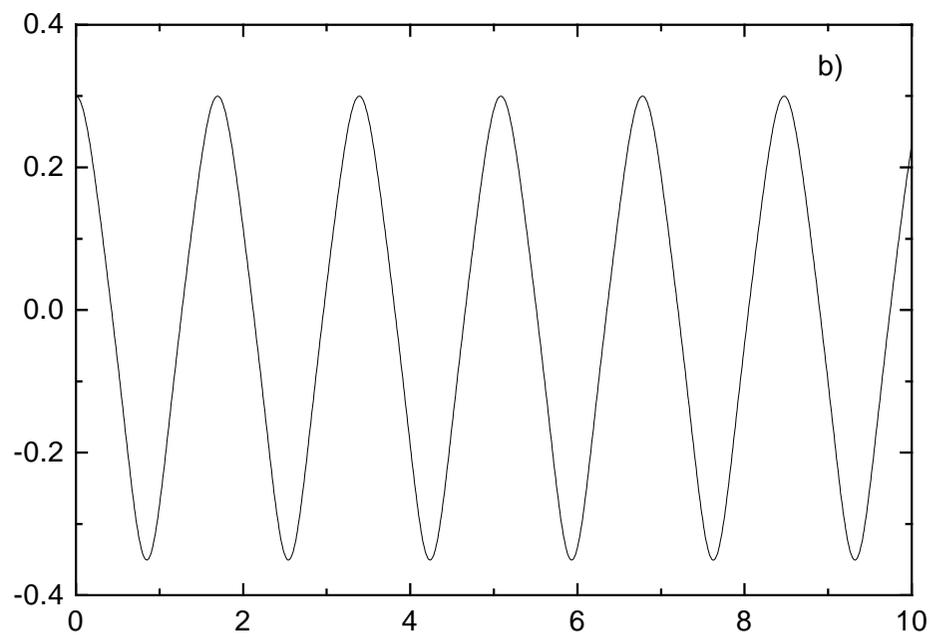
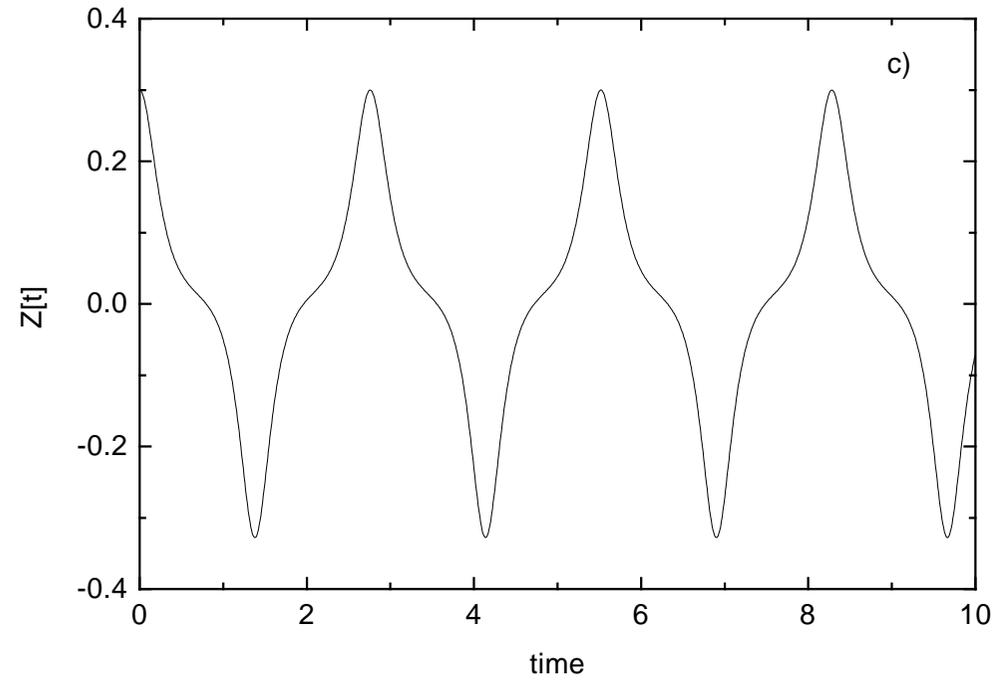
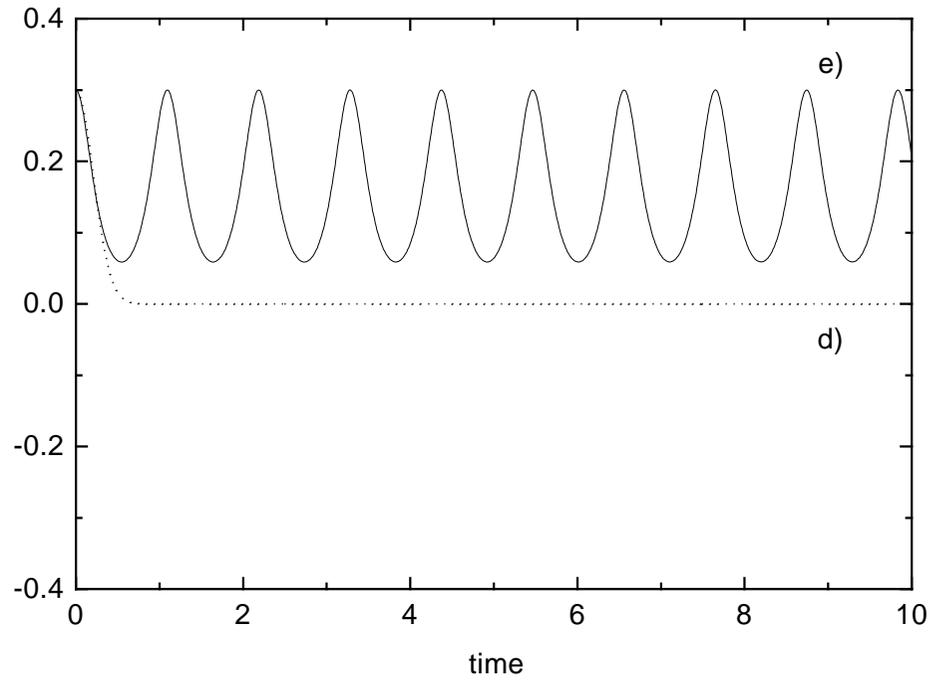
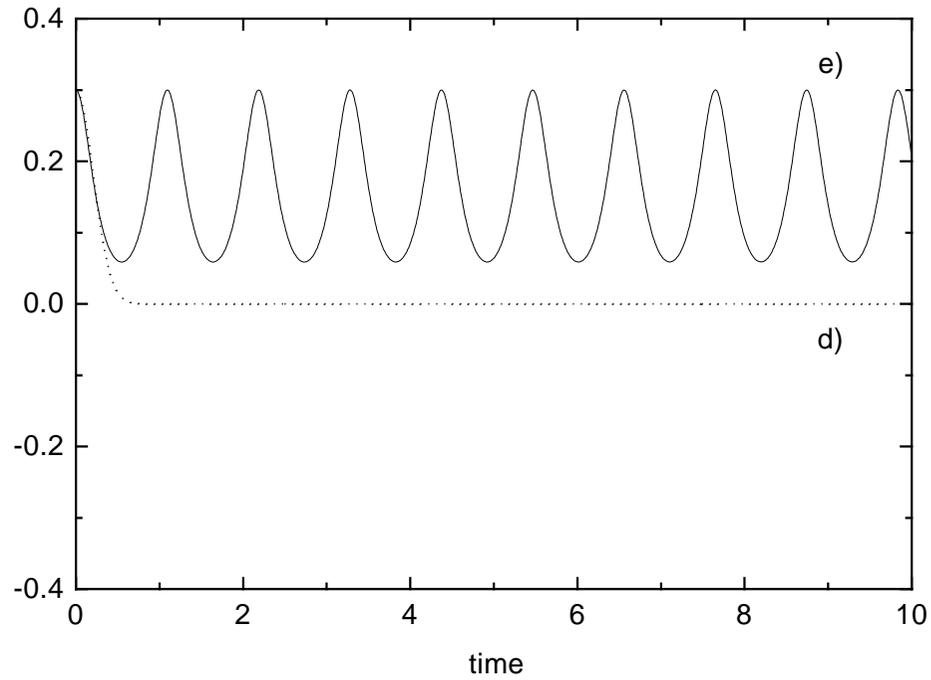
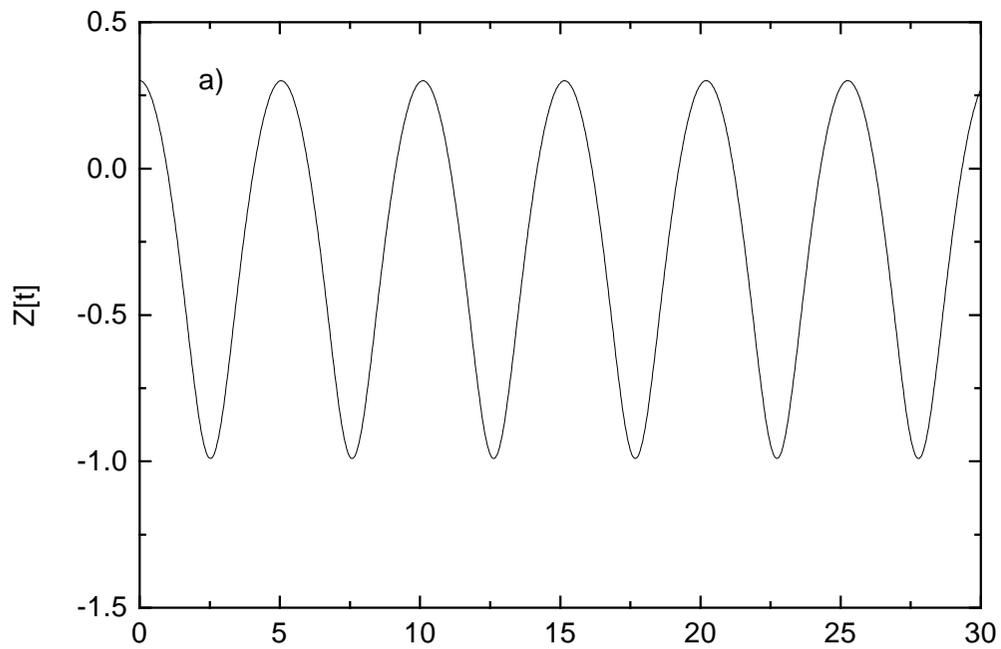
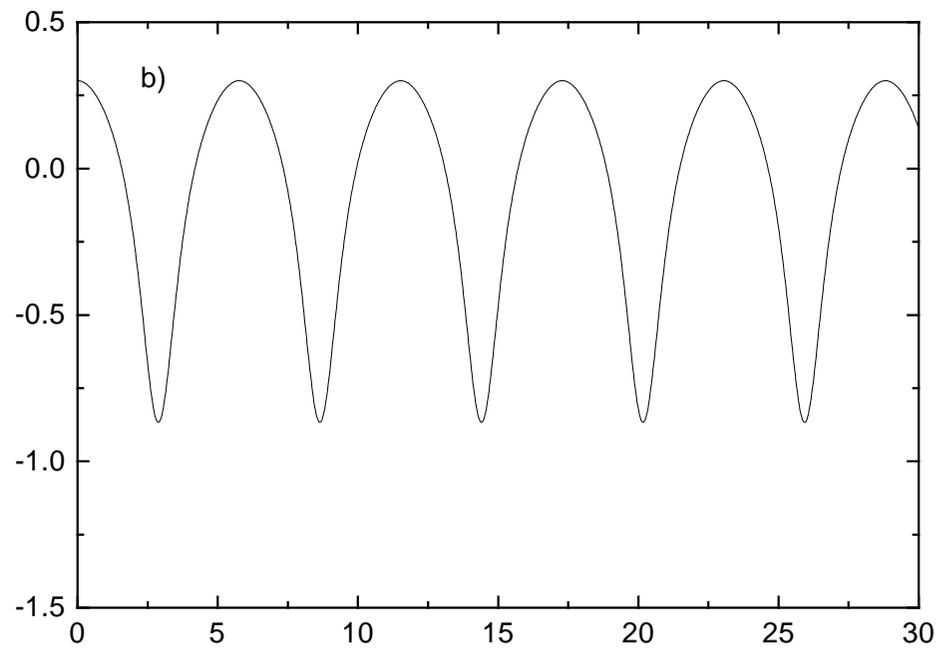
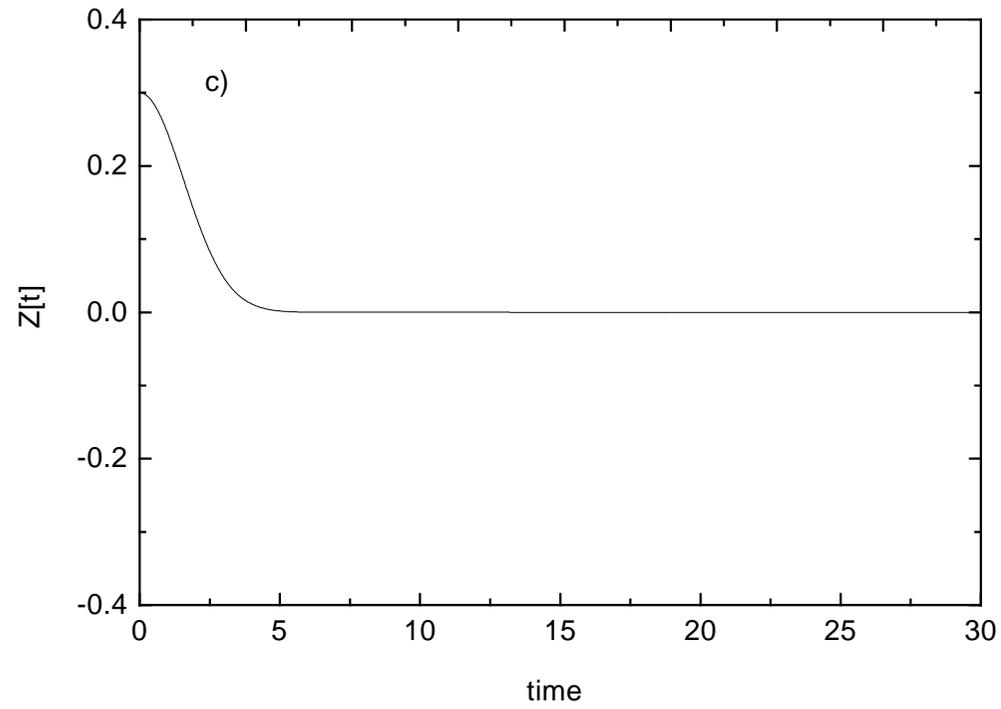
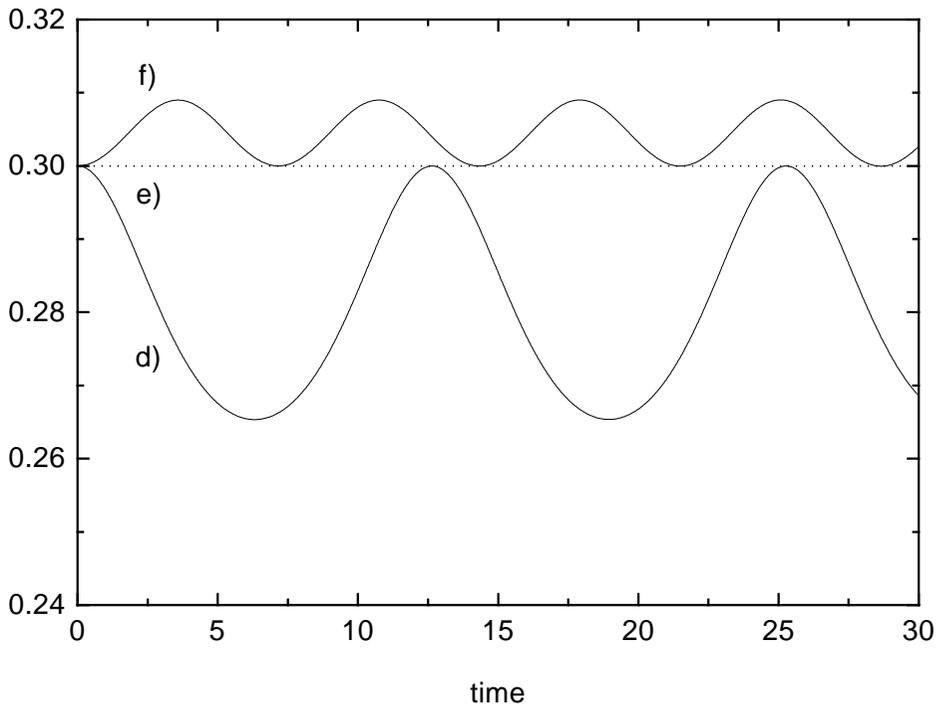
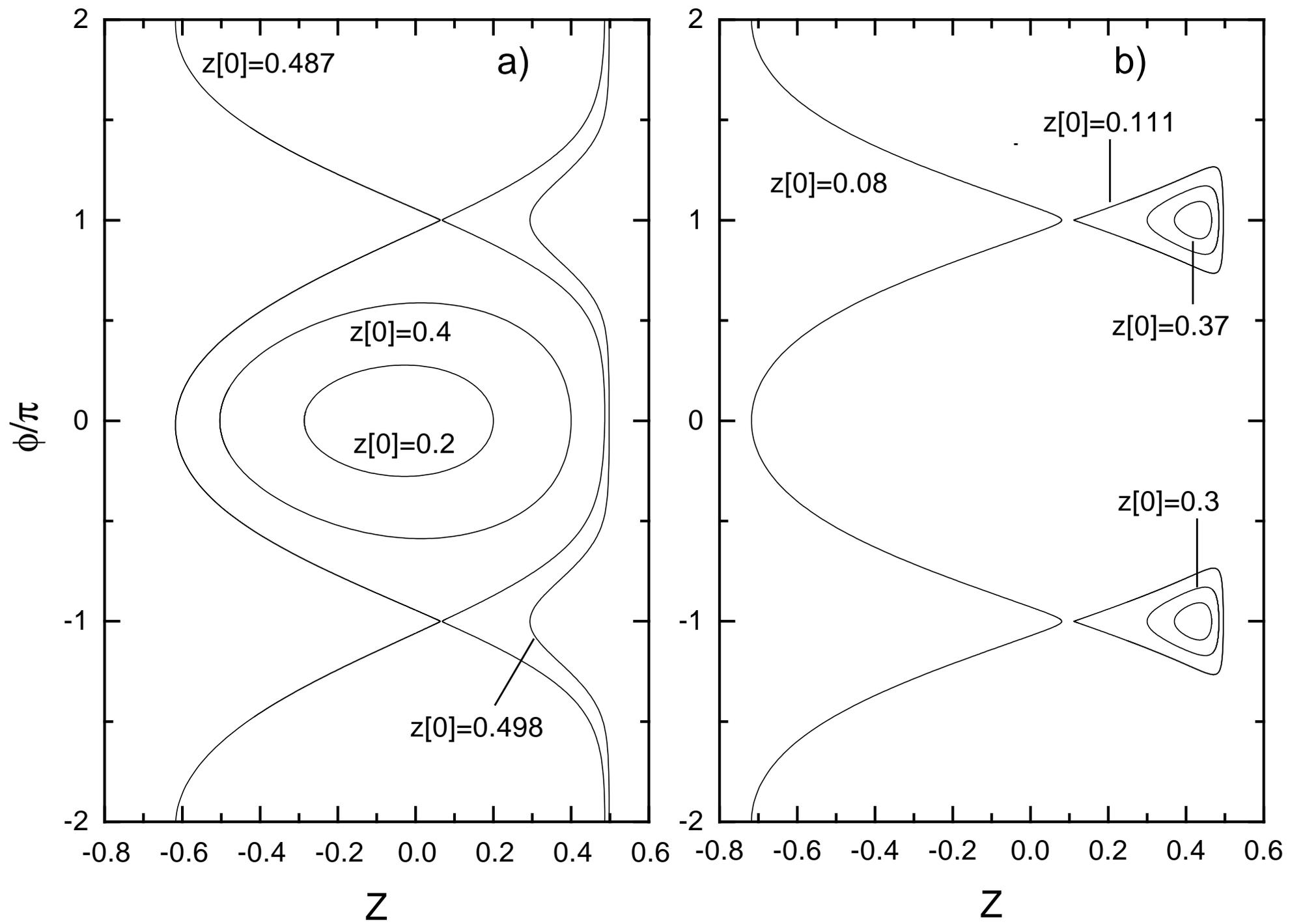
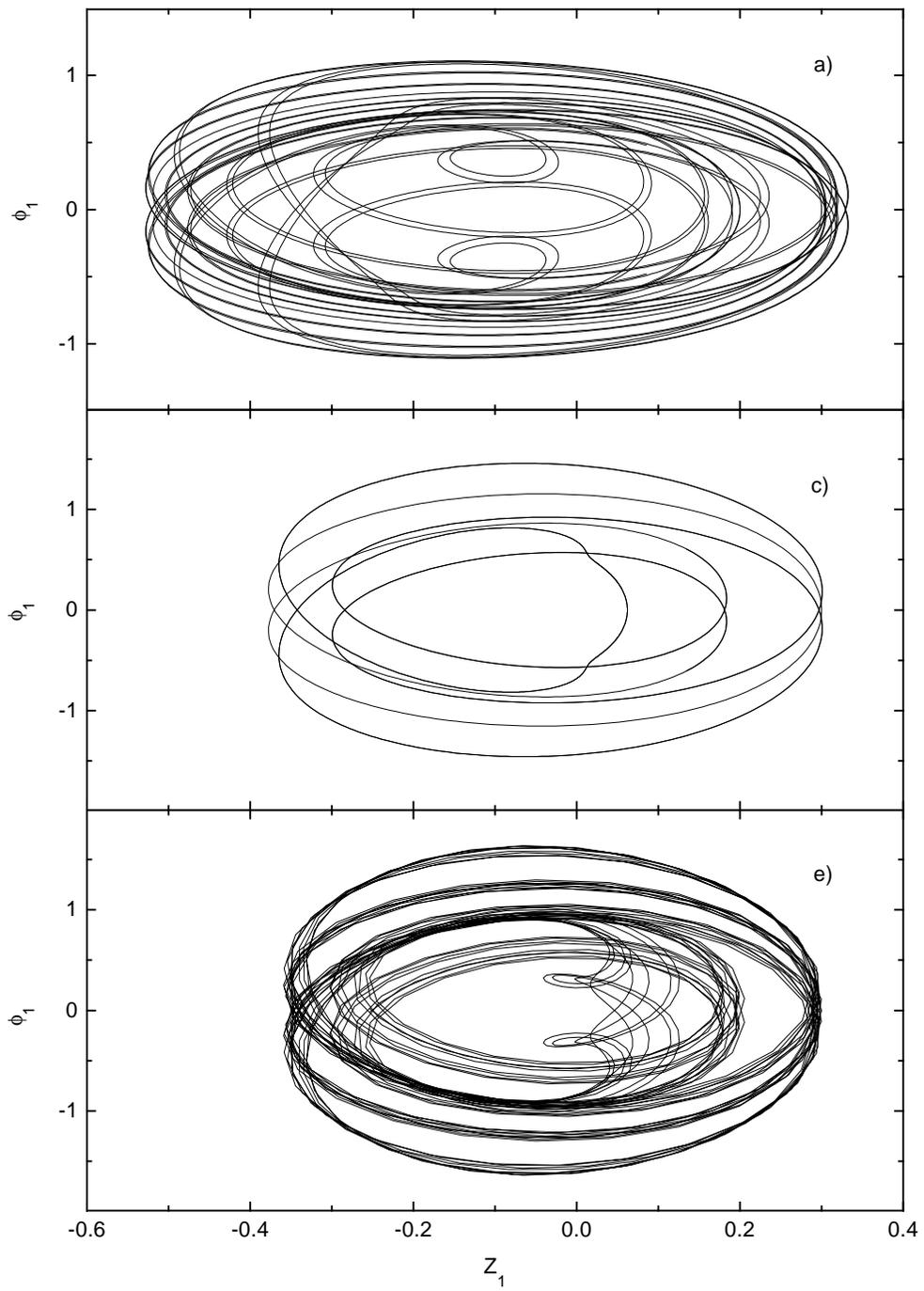
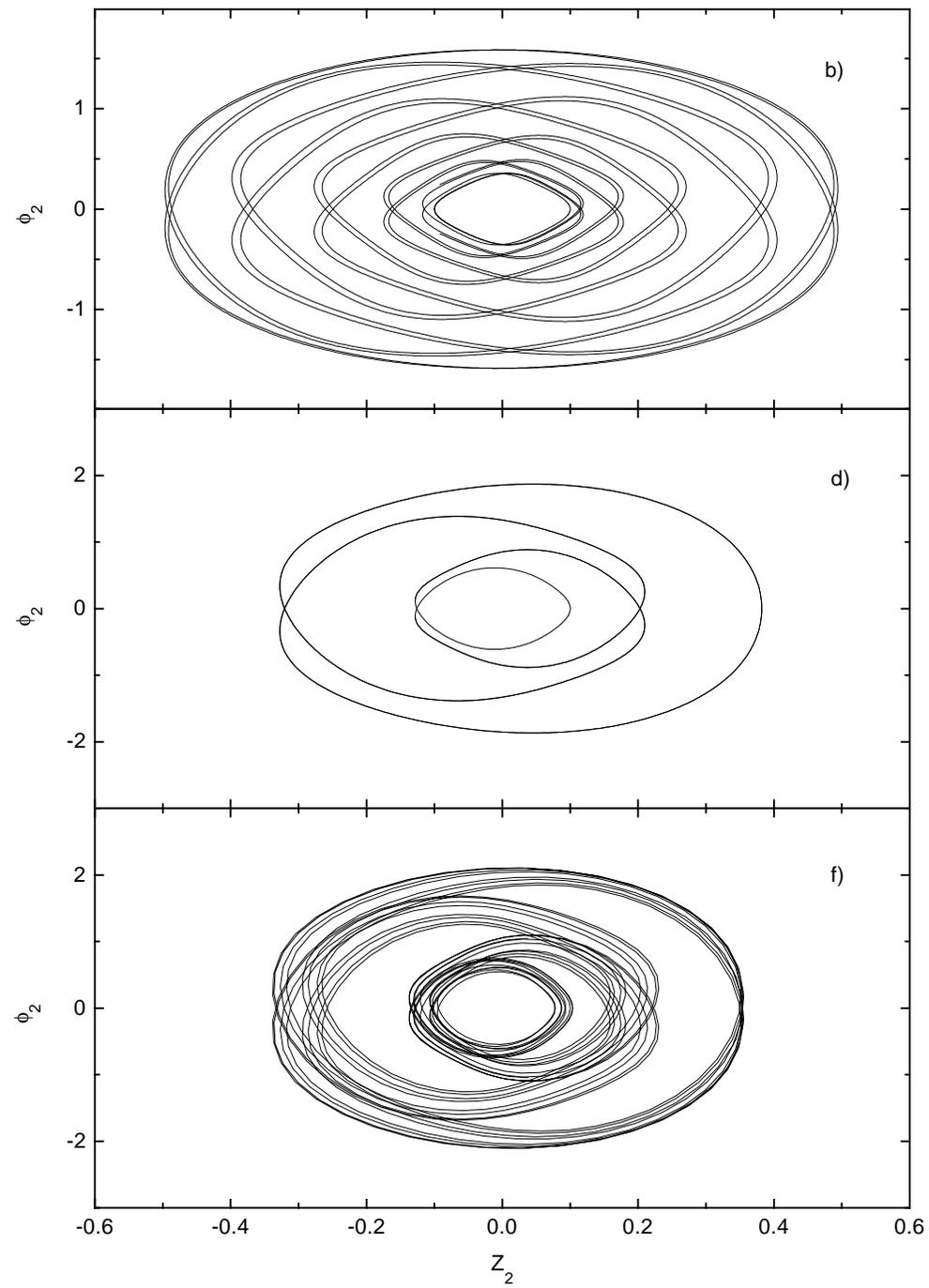
\end{document}